\documentclass[graybox]{svmult}
\usepackage{mathptmx}       
\usepackage{helvet}         
\usepackage{courier}        
\usepackage{makeidx}         
\usepackage{graphicx}        
\usepackage{multicol}        
\usepackage[bottom]{footmisc}
\usepackage{amsmath,amssymb,amsfonts}        
\usepackage{ stmaryrd } 

\usepackage{hyperref}

\newcommand{\cicy}[2]{\begin{matrix} #1\end{matrix}\!\left[\begin{matrix}#2 \end{matrix}\right]}

\newcommand{\sref}[1]{\S\ref{#1}}

\makeindex             

\begin{document}
\title*{A single point as a Calabi-Yau zerofold}
\author{Johanna Knapp and Joseph McGovern}
\institute{Johanna Knapp \at School of Mathematics and Statistics, University of Melbourne \\\email{johanna.knapp@unimelb.edu.au}
\and Joseph McGovern \at School of Mathematics and Statistics, University of Melbourne\\ \email{mcgovernjv@gmail.com}}
%
%
\maketitle

\abstract{We give a brief account of a non-abelian GLSM that describes a Calabi-Yau zerofold, in this case a single point, in the ``large volume'' phase. The other phase is non-regular, i.e.~there is no clear separation between the gauge and the matter sectors. Prepared for the proceedings of the MATRIX program ``The geometry of moduli spaces in string theory'' held in September 2024. 
} 

\section{Introduction}
The construction of the quintic threefold $\mathbb{P}^4[5]$ straightforwardly extends to complex dimension $n-1$ by way of $\mathbb{P}^{n}[n+1]$. This leads to the obvious realisation that $\mathbb{P}^1[2]$, i.e.~a generic quadric in $\mathbb{P}^1$, is a Calabi-Yau zerofold. In other words, two points are Calabi-Yau. This point has been made in \cite[\S7]{Candelas:2000fq}, where arithmetic methods were applied to CY varieties of various dimensions including $0$. While it may seem somewhat pointless to study zero-dimensional Calabi-Yaus, examples like this serve as toy models or motivation for certain constructions. See for instance \cite[\S 2.2]{Hori:2011pd} for a GLSM \cite{Witten:1993yc} account on $\mathbb{P}^1[2]$ as a preparatory example for the construction of a non-abelian duality in GLSMs. 

The point of this short note is two-fold. First, we want to point out the amusing fact that also a single point can be understood as a Calabi-Yau zerofold, in the sense that the axial R-symmetry of the associated GLSM is non-anomalous. We do so by giving a construction in terms of a one-parameter non-abelian GLSM. The construction is the zero-dimensional analogue of a non-abelian GLSM \cite{Hori:2011pd} whose two geometric phases are a pair of non-birational Calabi-Yaus first constructed by Hosono and Takagi \cite{Hosono:2011np,Hosono:2012hc,hosono2016double}. The model we present here is already implicit in \cite{Hori:2011pd} and other references below.

The second motivation is to produce a simple toy example for non-abelian GLSMs that have been termed ``irregular'' or ``non-regular''. First observed in \cite{Hori:2006dk} for non-abelian GLSMs associated to Grassmannians, these are GLSMs with phases where there is no scale separation between the gauge sector and the matter sector. A consequence is that field-theoretic methods to describe such phases, such as a Born-Oppenheimer approximation, do not apply and it becomes questionable whether it is possible at all to describe the low-energy behaviour of the GLSM in the phase in terms of a ``nice'' theory with an action such as a $(2,2)$ non-linear sigma model or a Landau-Ginzburg theory. A similar statement also holds for ``bad'' or ``pseudo-hybrid'' models \cite{Aspinwall:2009qy} which can also appear in abelian GLSMs, where a direct analysis of the low-energy physics is not straightforward. Since such phases of GLSMs are ubiquitous and may even be generic phases of GLSMs, they should be understood better. 

Recent accounts of non-regular GLSMs can be found in \cite{Guo:2025yed,Knapp:2025hnf}.  These models are fairly complicated, displaying many interrelated new phenomena, so it may be instructive to first study a simpler example. We hope that the model presented in this note can serve as a toy model for a non-regular GLSM. The non-regular model of \cite{Knapp:2025hnf}, studied by the authors, was related to a noncommutative resolution via topological string theory techniques developed in \cite{Schimannek:2021pau,Katz:2022lyl}. This demonstrates that such models can contain interesting mathematical structure. We present the non-regular model in this article as a simplest possible entry point in this program.

In the next section we introduce the model following \cite{Hori:2011pd}. We show that the ``large volume phase'' is a single point. We then show that the GLSM is non-regular by demonstrating that there is a Coulomb branch at infinite negative FI-parameter that coexists with a Higgs branch given by a determinantal quadric that cuts out two points in $\mathbb{P}^{1}$. We go on to consider the hemisphere and sphere partition functions, and study the divergent expressions that this produces in the strongly coupled phase. Moreover, we construct the mirror variety, which is a different description of a point, and compare with GLSM results, in particular the discriminant. In the final section we make suggestions for further research on this example.

\section{A non-abelian GLSM for a point}
We consider a non-abelian Calabi-Yau GLSM, with gauge group $G=(U(1)\times O(2))/{\{\pm 1,\pm{\bf 1}\}}\cong (U(1)\times U(1))\rtimes\mathbb{Z}_2$  and the following chiral matter content\footnote{By a common abuse of notation we use the same notation for superfields and their scalar components. The $\mathbb{Z}_{2}$ factor in the semidirect product exchanges the two $U(1)$ factors.}:
\begin{equation}\label{eq:ChargeMatrix}
      \begin{array}{c|cccccc|c}
        \phi&p^1&p^2&x_1&x_2&y_1&y_2&\mathrm{FI}\\
        \hline
        U(1)_1&-1&-1&1&1&0&0&\zeta\\
        U(1)_2&-1&-1&0&0&1&1&\zeta
        \end{array}
    \end{equation}
    The superpotential is
    \begin{equation}
      W=\sum_{i,j,k=1}^2S^{ij}_kp^kx_iy_j\equiv\sum_{i,j=1}^2S^{ij}(p)x_iy_j.
    \end{equation}
    The $\mathbb{Z}_2$ exchanges $x_i$ and $y_i$, and acts trivially on the $p$-fields. It also swaps the two vector multiplets. Therefore we can only have a single FI parameter $\zeta$. Due to the $\mathbb{Z}_2$-symmetry, the $2\times 2$ matrix $S^{ij}(p)$ is symmetric.

     The classical vacua are determined by the solutions of the D-term and F-term equations. The D-terms are
    \begin{equation}
      -\sum_{i=1}^2|p^i|^2+\sum_{i=1}^2|x_i|^2=\zeta,\qquad  -\sum_{i=1}^2|p^i|^2+\sum_{i=1}^2|y_i|^2=\zeta.
    \end{equation}
    From this, one concludes that the following deleted sets have to be excluded in the respective phases:
     \begin{equation}
      \label{htdeleted}
      F_{\zeta>0}=\{x_1=x_2=0\}\cup\{y_1=y_2=0\}, \qquad F_{\zeta<0}=\{p^1=p^2=0\}.
     \end{equation}
     The F-terms are given by $dW=0$.
     \subsection{Phases}
     The $\zeta>0$ phase is straightforward because the gauge symmetry is broken completely and the vacuum manifold is the Calabi-Yau
      \begin{equation}
      X=\{(x_i,y_i)\in(\mathbb{P}^1\times\mathbb{P}^1)/\mathbb{Z}_2|S^{ij}_k x_iy_j=0\}, \qquad k=1,2.
    \end{equation}
     This is indeed a single point: A complete intersection of two bilinears in $\mathbb{P}^1\times\mathbb{P}^1$ would give two points,
     \begin{equation}\begin{aligned}
\bigg(\;\frac{x_2}{x_1}\;,\;\frac{y_{2}}{y_{1}}\;\bigg)&=\bigg(\;\;\frac{\alpha\pm\sqrt{\nabla}}{2\beta}\quad,\quad\frac{\alpha\mp\sqrt{\nabla}}{2\beta}\;\;\bigg)\\[5pt]
\alpha&=S_{1}^{11}S_{2}^{22}-S_{2}^{11}S_{1}^{22},\qquad\qquad\beta=S_{2}^{12}S_{1}^{22}-S_{1}^{12}S_{2}^{22},\\[5pt]
\nabla&=\left(S^{11}_2 S^{22}_1-S^{11}_1 S^{22}_2\right){}^2+4 \left(S^{11}_2
   S^{12}_1-S^{11}_1 S^{12}_2\right) \left(S^{12}_1 S^{22}_2-S^{12}_2
   S^{22}_1\right).
     \end{aligned}\end{equation}
     However, these two points get identified by the $\mathbb{Z}_2$-action $(x_{1},x_{2})\leftrightarrow(y_{1},y_{2})$ that exchanges the two $\mathbb{P}^1$s, which amounts to flipping the sign in the $\pm$ above. This leaves us with just one point. 

     The $\zeta<0$ phase turns out to be more complicated due to strong coupling effects and the non-regularity of the model. The vacuum is given by $x_i=y_i=0$. It has an unbroken continuous symmetry $O(2)=U(1)\rtimes\mathbb{Z}_2$. The $p$-fields take VEVs in a $\mathbb{P}^1$. Turning on classical fluctuations of the $x$- and $y$-fields generates a potential of the form $W=S^{ij}(\langle p\rangle)x_{i}y_{j}$ and the symmetric $2\times 2$-matrix $S^{ij}(p)$ becomes a mass matrix. The $x$- and $y$-fields are massive except at the locus where the rank of $S(p)$ drops, i.e.~when the determinant vanishes:
     \begin{equation}\begin{aligned}
    \left(S^{11}_1 S^{22}_1-\left(S^{12}_1\right){}^2\right)(p^{1})^{2}+&\left(S^{11}_2 S^{22}_1+S^{11}_1 S^{22}_2-2 S^{12}_1 S^{12}_2\right)p^{1}p^{2}\\[3pt]&\qquad\qquad\;\;\;+\left(S^{11}_2 S^{22}_2-\left(S^{12}_2\right){}^2\right)(p^{2})^{2}=0.
     \end{aligned}\end{equation} Since the $p$-fields take values in $\mathbb{P}^1$, there can be a rank-$1$-locus, but there cannot be a rank-$0$-locus because for generic coefficients the only solution for $S^{ij}(p)=0$ is $p^1=p^2=0$ which is in the deleted set for this phase. So, in analogy to the Hosono-Takagi GLSM \cite{Hosono:2011np,Hori:2011pd}, one finds a symmetric determinantal variety
     \begin{equation}
       Y=\{(p^1,p^2)\in\mathbb{P}^1|\mathrm{rk}S(p)=1\}.
     \end{equation}
     The solution of the determinantal quadric gives two points rather than one:
     \begin{equation}\begin{aligned}
     \frac{p_{2}}{p_{1}}&=\frac{S^{11}_2 S^{22}_1+S^{11}_1 S^{22}_2-2 S^{12}_1 S^{12}_2\pm\sqrt{\Upsilon}}{2 \left(S^{12}_2\right){}^2-2S^{22}_2  S^{11}_2},\\[5pt]
     \Upsilon&=\left(S^{11}_2 S^{22}_1+S^{11}_1
   S^{22}_2-2 S^{12}_1 S^{12}_2\right){}^2-4 \bigg(\!\!\left(S^{12}_1\right){}^2-S^{11}_1
   S^{22}_1\bigg) \bigg(\!\!\left(S^{12}_2\right){}^2-S^{11}_2
   S^{22}_2\bigg).
     \end{aligned}\end{equation}
     If the $\zeta<0$ phase were just a non-linear sigma model with target $Y$, we could find a mismatch of Witten indices between the two phases, giving the first piece of evidence that our discussion is incomplete. In our argument we have implicitly made a Born-Oppenheimer approximation, assuming there is an energy scale where the gauge degrees of freedom decouple and the low-energy behaviour is determined by the matter sector. However, this is not true throughout the moduli space, the culprit being a Coulomb branch from the strongly coupled sector located at $\zeta\rightarrow-\infty$.

     \subsection{Coulomb branches}\label{sect:CoulombBranches}
     The Coulomb branch analysis for this model carries over from \cite{Hori:2006dk} which gives a much more general and comprehensive account and is, to our knowledge, the first reference to point out the existence of non-regular GLSMs. We start with discussing the Coulomb branch of the GLSM itself. The Coulomb branch loci are given in terms of the critical set of the effective superpotential
     \begin{equation}
       \label{weff}\begin{aligned}
             \mathcal{W}_{eff}=-t&(\sigma_1+\sigma_2)-2(-\sigma_1-\sigma_2)(\log(-\sigma_1-\sigma_2)-1)\\[2pt]&-2\sigma_1(\log(\sigma_1)-1)-2\sigma_2(\log(\sigma_2)-1),
     \end{aligned}\end{equation}
     where $\sigma_{1,2}$ are the vector multiplets. The critical locus is at
    \begin{equation}
    \label{coulomb}
      e^{-t}=\frac{\sigma_1^2}{(\sigma_1+\sigma_2)^2}=\frac{\sigma_2^2}{(\sigma_1+\sigma_2)^2}.
    \end{equation}
    From this it follows that $z^2=\frac{\sigma_2^2}{\sigma_1^2}=1$. This gives a Coulomb branch at $\varphi\equiv e^{-t}=\frac{1}{4}$ from the locus $\sigma_1=\sigma_2$. Since this locus is fixed by the $\mathbb{Z}_2$ Weyl group of the non-abelian gauge group, there are actually two Coulomb branches at this point \cite{Hori:2011pd}. We expect that this is related to the fact that the quotient action defining the mirror $M$ in \sref{sect:MirrorSymmetry} does not act freely for $\varphi=\frac{1}{4}$. The equations (\ref{coulomb}) also have a solution for $\sigma_1+\sigma_2=0$, indicating another Coulomb branch at $\zeta\rightarrow-\infty$. However, since the $p$-fields are massless at this locus, it is not allowed to integrate them out, making (\ref{weff}) invalid.

    The locus $\sigma_1+\sigma_2=0$, however, makes a reappearance in the strongly coupled phase and eventually does lead to a Coulomb branch at $\zeta\rightarrow-\infty$. The condition parametrises the unbroken gauge symmetry in the strongly coupled phase. As was pointed out in \cite[\S 4.2]{Hori:2011pd}, Hosono-Takagi-type GLSMs with $N$ fundamentals\footnote{In our model, the fundamentals consist of linear combinations of $x$- and $y$-fields.} and gauge group $O(k)$ are non-regular when $N-k$ is even, meaning that there is a Coulomb branch for any negative value of the FI parameter. This implies that the gauge degrees of freedom do not decouple. The condition for non-regularity is satisfied in our case, which is $N=k=2$. It has been further argued in \cite[\S 4.4]{Hori:2006dk} in an analysis of models with unbroken $SU(k)$ that this Coulomb branch is lifted almost everywhere by dynamically creating masses for the $\sigma$-fields. The discussion carries over almost word by word to our case. We choose a parametrisation  \begin{equation}
      \sigma_1=\sigma-\sigma_0, \qquad \sigma_2=-\sigma,
    \end{equation}
    where $\sigma_0$ is associated to the $U(1)$ that is Higgsed in the strongly coupled phase. Inserting this into (\ref{weff}) and integrating out $\sigma_0$ leads to
    \begin{equation}
      \sigma_0=\frac{1}{1+e^{-\frac{t}{2}}}\sigma=:f(e^{-t})\sigma
    \end{equation}
    with
    \begin{equation}
      t+2\log(\sigma(1-f))-2\log(f\sigma)=0.
    \end{equation}
    Taking this into account and eliminating $\sigma_0$ from $\mathcal{W}_{eff}$ gives
    \begin{equation}
      \mathcal{W}_{eff}(\sigma)=2\sigma\left(\log\frac{-1}{(1-f)}\right).
    \end{equation}
    This has a critical locus at $f=2$ which is the same as $e^{-t}=\frac{1}{4}$ which is the Coulomb branch locus we have already found. This means there is no Coulomb branch in the $\zeta\ll0$ phase, except at $\zeta\rightarrow -\infty$. This poses a conundrum. We have one point in one phase, and two points plus a Coulomb branch in the other, with an unbroken nonabelian continuous gauge symmetry. It is not obvious how these two phases can be equivalent in a way phases of other GLSMs are. We do not know how to resolve this issue but make some comments in \sref{sec-beyond}. 
    \subsection{Structure sheaf and hemisphere partition function}
    Coming back to the $\zeta\gg0$ phase, we can make a few further statements about our Calabi-Yau zerofold by computing the hemisphere partition function for the structure sheaf. The hemisphere partition function \cite{Sugishita:2013jca,Honda:2013uca,Hori:2013ika} for the model is\footnote{We have fixed the R-charge ambiguity to match with the $\zeta\gg0$ phase.}
    \begin{equation}
  Z_{D^2}=\frac{C}{(2\pi)^{2}}\int d^2\sigma \Gamma(1-\text{i}\sigma_2-\text{i}\sigma_2)^2\Gamma(\text{i}\sigma_1)^2\Gamma(\text{i}\sigma_2)^2e^{\text{i}t(\sigma_1+\sigma_2)}f_{\mathcal{B}}(\sigma),
    \end{equation}
    where the brane factor $f_{\mathcal{B}}=\mathrm{Tr}_Me^{\text{i}\pi r_*}e^{2\pi\rho(\sigma)}$ is the character of the representations $r_*$ and $\rho$ of the vector R- and gauge symmetries on the Chan-Paton space $M$ of the GLSM brane $\mathcal{B}$ \cite{Herbst:2008jq,Hori:2013ika}. To evaluate the partition function in the $\zeta\gg0$ phase we close the integration contour such that the poles of $\Gamma(\sigma_{1,2})$ are enclosed, i.e.~we take $\text{i}\sigma_i=- k_i+\varepsilon_i$ for $k_i\in\mathbb{Z}_{\geq 0}$. This guarantees a convergent result in the phase. The integral can be rewritten as
    \begin{align}
      \label{zd2lv}
  Z_{D^2}^{\zeta\gg 0}=\frac{C}{(2\pi\text{i})^2}\sum_{k_1,k_2\geq 0}\oint d^2\varepsilon & \frac{\pi^4}{\sin^2\pi\varepsilon_1\sin^2\pi\varepsilon_2}\nonumber\\
  &\cdot\frac{\Gamma(1+k_1+k_2-\varepsilon_1-\varepsilon_2)^2}{\Gamma(1+k_1-\varepsilon_1)^2\Gamma(1+k_2-\varepsilon_2)^2}e^{-t(k_1+k_2-\varepsilon_1-\varepsilon_2)}f_{\mathcal{B}}.
  \end{align}
    The brane that describes the structure sheaf is given by the following matrix factorisation of the superpotential:
    \begin{equation}
      Q=\sum_{k=1}^2p^k\eta_k+S^{ij}_kx_iy_j\overline{\eta}_k,
    \end{equation}
    where $\eta_k,\overline{\eta}_k$ satisfy a Clifford algebra. We can associate a GLSM B-brane $\mathcal{B}$ to this which has the brane factor
    \begin{equation}
       f_{\mathcal{B}}=(1-e^{2\pi(\sigma_1+\sigma_2)})^2.
      \end{equation}
    Inserting this into (\ref{zd2lv}), we find
    \begin{equation}
  Z_{D^2}^{\zeta\gg0}(\mathcal{B})=2C\varpi_0,
    \end{equation}
    where\footnote{Recall that $\varphi=e^{-t}$.}
      \begin{align}\label{eq:GLSMperiod}
      \varpi_0=&\sum_{k_1,k_2=0}^{\infty}\frac{\Gamma(1+k_1+k_2)^2}{\Gamma(1+k_1)^2\Gamma(1+k_2)^2}\varphi^{k_1+k_2}\nonumber\\
  =&1+2 \varphi+6 \varphi^2+20 \varphi^3+70 \varphi^4+252 \varphi^5+O\left(\varphi^6\right)\nonumber\\
  =&\frac{1}{\sqrt{1-4\varphi}}.
  \end{align}
      This is annihilated by the Picard-Fuchs operator
\begin{equation}\label{eq:PFoperator}\begin{aligned}
\mathcal{L}&=\;\;(1-4\varphi)\theta-2\varphi\;\;=\;\;\theta-2\varphi(2\theta+1),\qquad\theta=\varphi\frac{\text{d}}{\text{d}\varphi}.
\end{aligned}\end{equation}
The Riemann symbol for $\mathcal{L}$ is 
\begin{equation}
\left\{\begin{matrix}0&\phantom{-}\frac{1}{4}&\phantom{-}\infty\\[4pt]0&-\frac{1}{2}&\phantom{-}\frac{1}{2}\end{matrix}\right\}.
\end{equation}
The period $\varpi_{0}(\varphi)$ is the same as for the mirror of $\mathbb{P}^1[2]$ (see for instance \cite[\S7]{Candelas:2000fq}). 

We can also obtain $\varpi_0$ by evaluating the hemisphere partition function for the associated $U(1)$ GLSM, whose $\zeta\gg0$ phase is $\mathbb{P}^{1}[2]$, which is easily adapted from the quintic case discussed in \cite{Herbst:2008jq,Hori:2013ika}. This returns $\varpi_{0}=\sum_{m\geq0}\frac{(2m)!}{m!^2}\varphi^{m}=\frac{1}{\sqrt{1-4\varphi}}$. Distinguishing two points from one point\footnote{Note that there is also a $U(1)\times\mathbb{Z}_{2}$ model to describe a point, with $\zeta\gg0$ phase $\mathbb{P}^{1}[2]/\mathbb{Z}_{2}$, but that is not the object of discussion here.} hence boils down to the choice of the normalisation constant $C$. We propose that the correct choice for this model is $C=\frac{1}{2}$, where we interpret the factor $2$ in the denominator as the rank of the Weyl group, consistent with \cite{Hori:2013ika}. In contrast, for the $\mathbb{P}^1[2]$-model we should choose the normalisation in such a way that the central charge of the structure sheaf is $2\varpi_0$. 
      
      Finally, we can consider the $\zeta\ll0$ phase where we find further issues related to the non-regularity of the model. A prescription for evaluating the sphere partition function in the $U(2)$ GLSM describing the R{\o}dland model \cite{rodland2000pfaffian,Hori:2006dk} has been given in \cite{Jockers:2012dk}. We apply this to our model and take the poles
      \begin{equation}
      \label{poles}
        1-\text{i}\sigma_1-\text{i}\sigma_2=-n_p,\qquad \text{i}\sigma_1=-n_1, \qquad n_1,n_p\in\mathbb{Z}_{\geq 0}.
        \end{equation}
Then the integral can be written as
\begin{align}
  Z_{D^2}^{\zeta\ll 0}=-\frac{C}{(2\pi\text{i})^2}\sum_{n_1,n_p=0}^{\infty}\oint & d\varepsilon_1 d\varepsilon_p \frac{\Gamma(1+n_1+n_p-\varepsilon_1-\varepsilon_p)^2}{\Gamma(1+n_1-\varepsilon_1)^2\Gamma(1+n_p-\varepsilon_p)^2}\nonumber\\
  &\cdot\frac{\pi^4}{\sin^2\pi(-n_1+\varepsilon_1)\sin^2\pi(-n_p-\varepsilon_p)} e^{t(1+n_p-\varepsilon_p)}\nonumber\\
  &\cdot f_{\mathcal{B}}(\text{i}(n_1-\varepsilon_1),-i(1+n_1+n_p-\varepsilon_1-\varepsilon_p)).
\end{align}
The sum in $n_1$ does not converge for any brane factor\footnote{The dependence on $n_1,n_p$ comes from $e^{2\pi\rho(\sigma)}$ which produces terms of the form $e^{2\pi(a\sigma_1+b\sigma_2)}$ for $a,b\in\mathbb{Z}$. Inserting (\ref{poles}), the $n_1,n_p$-dependence will always drop out.}. In contrast to the odd-dimensional case, the $\sin$-term gives a $(-1)^{2n_1}=1$ rather than alternating signs $(-1)^{n_1}$. Hence, the convergence behaviour is worse compared to the threefold and elliptic curve case, but also occurs in the two-dimensional case which is also non-regular. We should not interpret this as the result being divergent. Rather, it tells us that the integration contour we have chosen is not valid for non-regular models. It would be interesting to show if a contour exists that gives a convergent result.

\subsection{Sphere partition function}
The GLSM can also be placed on a sphere. As shown in \cite{Jockers:2012dk}, the partition function of this theory, which is not topological, can be used to obtain period functions and genus-0 invariants of Calabi-Yau threefolds. Studying this partition function for our model will allow us to further probe the non-regular $\zeta\ll0$ phase, and gives a complementary view on divergences that arise.

After assigning $U(1)_{R}$ charges $2q$ to the $x_{i},y_{i}$ fields and $2-4q$ to the $p_{i}$ fields, we can compute the sphere partition function for our model following \cite{Jockers:2012dk}. With the GLSM data for our model understood, this takes the form\footnote{Note $\zeta_{\text{here}}=\frac{1}{2\pi}r_{\text{there}}$.}
\begin{equation}\label{eq:ZS2general}
Z_{S^{2}}=\frac{1}{2}\sum_{m_{i}\in\mathbb{Z}}\text{e}^{-\text{i}\theta(m_{1}+m_{2})}\int_{-\infty}^{\infty}\int_{-\infty}^{\infty}\frac{\text{d}\sigma_{1}\text{d}\sigma_{2}}{(2\pi)^{2}}\text{e}^{-2\text{i}\zeta(\sigma_{1}+\sigma_{2})}Z_{\text{matter}}(\sigma_{i},m_{i})Z_{\text{gauge}}(\sigma_{i},m_{i}),
\end{equation}
where $Z_{\text{matter}}=\frac{\Gamma(q-\text{i}\sigma_{1}-\frac{m_{1}}{2})^{2}}{\Gamma(1-q+\text{i}\sigma_{1}-\frac{m_{1}}{2})^{2}}\frac{\Gamma(q-\text{i}\sigma_{2}-\frac{m_{2}}{2})^{2}}{\Gamma(1-q+\text{i}\sigma_{2}-\frac{m_{2}}{2})^{2}}$ and $Z_{\text{gauge}}=\frac{\Gamma(1-2q+\text{i}\sigma_{1}+\text{i}\sigma_{2}+\frac{m_{1}+m_{2}}{2})^{2}}{\Gamma(2q-\text{i}\sigma_{1}-\text{i}\sigma_{2}+\frac{m_{1}+m_{2}}{2})^{2}}$ are prescribed by the 2d $\mathcal{N}=(2,2)$ supersymmetric localisation results of \cite{Benini:2012ui,Doroud:2012xw}. The prefactor of $\frac{1}{2}$ is taken because our Weyl group is $\mathbb{Z}_{2}$.

To evaluate $Z_{S^{2}}$ in the $\zeta\gg0$ phase we close both $\sigma_{i}$ contours in their lower half planes. The integral is performed by residues, with contributing poles at $\sigma_{i}=-\text{i}q+\text{i}(\frac{m_{i}}{2}-k_{i})$, with $k_{i}\in\mathbb{Z}_{\geq0}$. Following \cite{Jockers:2012dk}, we compute
\begin{equation}\begin{aligned}
Z_{S^{2}}^{\zeta\gg0}&=\frac{(\varphi\overline{\varphi})^{2q}}{2}\oint\frac{\text{d}\epsilon_{1}\text{d}\epsilon_{2}}{(2\pi\text{i})^{2}}(\varphi\overline{\varphi})^{-\epsilon_{1}-\epsilon_{2}}\left(\frac{\pi\sin(\pi(\epsilon_{1}+\epsilon_{2}))}{\sin(\pi\epsilon_{1})\sin(\pi\epsilon_{2})}\right)^{2}\\&\qquad\qquad\qquad\qquad\bigg|\sum_{n_{i}\geq0}\frac{\Gamma(1+n_{1}+n_{2}-\epsilon_{1}-\epsilon_{2})^{2}}{\Gamma(1+n_{1}-\epsilon_{1})^{2}\Gamma(1+n_{2}-\epsilon_{2})^{2}}\varphi^{n_{1}+n_{2}}\bigg|^{2}\\[5pt]&=(\varphi\overline{\varphi})^{2q}\varpi_{0}(\varphi)\overline{\varpi_{0}(\varphi)},
\end{aligned}\end{equation}
where, as in \cite{Jockers:2012dk}, the complex conjugation does not act of the variables $\epsilon_{i}$, and the contours in $\epsilon_{i}$-space are small circles around 0. We reiterate that $\varphi=\text{e}^{-t}=\text{e}^{-\zeta+\text{i}\theta}$. 

In the phase $\zeta\ll0$, we close both $\sigma_{i}$ contours in the upper half plane, with poles at $\sigma_{1}+\sigma_{2}=-2\text{i}q+\text{i}(\frac{m_{1}+m_{2}}{2})+k_{p}$, $\sigma_{1}=-\text{i}q+\text{i}(1+\frac{m_{1}}{2}+k_{1}+k_{p})$. We obtain
\begin{equation}\begin{aligned}
Z_{S^{2}}^{\zeta\gg0}&=\frac{(\varphi\overline{\varphi})^{2q-1}}{2}\oint\frac{\text{d}\epsilon_{1}\text{d}\epsilon_{p}}{(2\pi\text{i})^{2}}(\varphi\overline{\varphi})^{-\epsilon_{p}}\left(\frac{\pi\sin(\pi(\epsilon_{1}+\epsilon_{p}))}{\sin(\pi\epsilon_{1})\sin(\pi\epsilon_{p})}\right)^{2}\\&\qquad\qquad\qquad\qquad\bigg|\sum_{n_{i}\geq0}\frac{\Gamma(1+n_{1}+n_{p}+\epsilon_{1}+\epsilon_{p})^{2}}{\Gamma(1+n_{1}+\epsilon_{1})^{2}\Gamma(1+n_{p}+\epsilon_{p})^{2}}\varphi^{-n_{p}}\bigg|^{2}\\[5pt]&=(\varphi\overline{\varphi})^{2q-1}\bigg|\sum_{n_{i}\geq0}\frac{\Gamma(1+n_{1}+n_{p})^{2}}{\Gamma(1+n_{1})^{2}\Gamma(1+n_{p})^{2}}\varphi^{-n_{p}}\bigg|^{2}.
\end{aligned}\end{equation}
We have taken residues to arrive at the final expression above, however, the series is divergent. Unlike in the $\zeta\gg0$ phase and consistent with the hemisphere partition function, the sum over $n_{1}$ at each fixed $n_{p}$ is divergent for all $\varphi$.

This is reminiscent of the situation concerning the Pfaffian phase of the R{\o}dland model as studied by \cite{Jockers:2012dk}. There, a convergence factor $\text{e}^{-\delta n_{1}}$ was incorporated in the sum, so ensuring convergence of the sum for positive $\delta$, and taking the limit $\delta\rightarrow0^{+}$. However, for this to work one relies on the presence of a $(-1)^{n_{1}}$ in their summand. There is no such $(-1)^{n_{1}}$ in our summand. As already observed for the hemisphere,  more severe divergences seem to be a feature of non-regular models.

Let us consider more closely the problem of regulating the sum
\begin{equation}
S(\varphi)=\sum_{n_{1},n_{p}\geq0}\frac{\Gamma(1+n_{1}+n_{p})^{2}}{\Gamma(1+n_{1})^{2}\Gamma(1+n_{p})^{2}}\varphi^{-n_{p}}.
\end{equation}
If we rotate the $\sigma_{1}$-contour counter-clockwise by a small angle $\delta$, so include a convergence factor $\text{e}^{\text{i}\delta \sigma_{1}}$ in \eqref{eq:ZS2general}, then we instead obtain
\begin{equation}\begin{aligned}
S_{\delta}(\varphi)&= \sum_{n_{1},n_{p}\geq0}\frac{\Gamma(1+n_{1}+n_{p})^{2}}{\Gamma(1+n_{1})^{2}\Gamma(1+n_{p})^{2}}\text{e}^{-\delta n_{1}}\varphi^{-n_{p}}\\[5pt] &=\sum_{n_{p}\geq0}{}_{2}F_{1}(1+n_{p},1+n_{p};1;\text{e}^{-\delta})\varphi^{-n_{p}}. 
\end{aligned}\end{equation}
The hypergeometric functions account for the divergences we described. Typically one can analytically continue a series ${}_{p+1}F_{p}$ to argument $-1$, which is why the models with such $(-1)^{n_{1}}$ factors in their summands are not problematic. See for instance the expression in \cite[eq. 2-5.4]{Hosono:2011np}, which could also be obtained by a GLSM computation like the ones we are describing. But an argument of $1$ in the hypergeometric function is more problematic: there can be a pole. Two useful identities are ${}_{2}F_{1}(a,b;c;z)=(1-z)^{c-a-b}{}_{2}F_{1}(c-a,c-b;c;z)$ for $|\text{Arg}(1-z)|<\pi$, and ${}_{2}F_{1}(a,b;c;1)=\frac{\Gamma(c)\Gamma(c-a-b)}{\Gamma(c-a)\Gamma(c-b)}$ for $\text{Re}(c-a-b)>0$ \cite[\S15]{NIST:DLMF}. These imply
\begin{equation}\label{eq:divergence}
{}_{2}F_{1}(1+n_{p},1+n_{p};1;\text{e}^{-\delta})=\frac{(2n_{p})!}{n_{p}!^{2}}\frac{1}{\delta^{2n_{p}+1}}\big(\,1+O(\delta)\,\big).
\end{equation}
Based on this equation, we will attempt a rescaling. First we multiply $S_{\delta}$ by $\delta$, and then we make a rescaling of our algebraic coordinate, $\varphi=\delta^{-2}\widetilde{\varphi}$. Only then do we take $\delta\rightarrow0^{+}$, finding
\begin{equation}\begin{aligned}
\delta S_{\delta}(\delta^{-2}\widetilde{\varphi})\bigg\vert_{\delta\rightarrow0^{+}}&=\sum_{n_{p}\geq0}\left[\delta^{1+2n_{p}}\,{}_{2}F_{1}(1+n_{p},1+n_{p};1;\text{e}^{-\delta})\right]_{\delta\rightarrow0^{+}}\widetilde{\varphi}^{-n_{p}}\\[5pt]
&=\sum_{n_{p}\geq0}\frac{(2n_{p})!}{n_{p}!^{2}}\widetilde{\varphi}^{-n_{p}}=\frac{1}{\sqrt{1-4\widetilde{\varphi}^{-1}}}=\varpi_{0}(\widetilde{\varphi}^{-1}),
\end{aligned}\end{equation}
where we employ \eqref{eq:divergence}. The appearance of the period function from the $\zeta\gg0$ phase is reassuring, although we do not have a full interpretation of this manipulation. It would be interesting to better understand the scaling $\varphi=\delta^{-2}\widetilde{\varphi}$, which as a modification of the FI-parameter $\zeta$ takes the form
\begin{equation}\label{eq:ZetaLog}
\zeta=\widetilde{\zeta}+2\log(\delta).
\end{equation}

Perhaps a better way to obtain the period in the $\zeta\ll0$ phase is to forego the sphere partition function, and instead obtain an expansion about $\phi=-\frac{1}{16\varphi^{2}}=0$ by using the closed form $\varpi_{0}(\varphi)=\frac{1}{\sqrt{1-4\varphi}}$. This gives
\begin{equation}\begin{aligned}
\varpi_{0}(\phi)&=\frac{2\phi}{\sqrt{1+4\phi^{2}}}=2\phi\sum_{k=0}^{\infty}\frac{(2k)!}{k!^{2}}(-1)^{k}\phi^{2k}
\\[5pt]&=2\phi(1 - 2\phi^2 + 6\phi^4 - 20\phi^6 + 70\phi^8 - 
 252\phi^{10} + 924\phi^{12}+...).
\end{aligned}\end{equation}
The number $-1/16$ in our change of variables is chosen such that we get a series with integral coefficients. Our exponent $-2$ in $\phi=-2^{-4}\varphi^{-2}$ is chosen because the Riemann symbol of $\mathcal{L}$ has an entry $1/2$ at the singularity $\infty$, see the parallel discussion in \cite{Katz:2022lyl} concerning their change of variable. In the $\phi$-plane, there are two singularities at $\phi=\pm\frac{\text{i}}{2}$ equidistant from the singularity $\phi=0$ about which we are expanding. This seems to be a common property of noncommutative resolution phases of Calabi-Yau moduli spaces, but further study is needed to relate such structures to our model. 

     \subsection{Mirror Symmetry}\label{sect:MirrorSymmetry}
Here we describe the mirror construction using toric methods. The previous Coulomb branch computation in the GLSM provides the discriminant of the mirror variety. Although the mirror variety is a single point that does not obviously degenerate in any way, we are able to relate the discriminant to the failure of a quotient action to be fixed-point free.
     
     We have described the $\zeta>0$ geometry $X$ as a $\mathbb{Z}_{2}$ quotient of the variety
     \begin{equation}\label{eq:cicy2pt}
\widetilde{X}\;\;\;\cong\;\;\;\cicy{\mathbb{P}^{1}\\\mathbb{P}^{1}}{1&1\\1&1},
     \end{equation}
where we use the CICY notation \cite{Candelas:1987kf,Green:1986ck}. We can use the procedure described in \cite[\S3]{Hosono:1994ax} to obtain the mirror variety $\widetilde{M}$ of $\widetilde{X}$ as the locus
\begin{equation}
\widetilde{M}:\qquad U+V=1,\quad\quad \frac{\varphi_{1}}{U}+\frac{\varphi_{2}}{V}=1
\end{equation}
in the torus $(\mathbb{C}^*)^{2}$ with coordinates $U,V$. This is the zero-dimensional analogue of the threefold families studied by Hulek and Verrill \cite{hulek2005modularity}. This mirror variety $\widetilde{M}$ is also two points, given by \begin{equation}\label{eq:MirrorSolution}
(U,V)=\bigg(\frac{1+\varphi_{1}-\varphi_{2}\pm\sqrt{\Delta}}{2},\frac{1-\varphi_{1}+\varphi_{2}\mp\sqrt{\Delta}}{2}\bigg).
\end{equation}The discriminant is
\begin{equation}\begin{aligned}
\Delta(\varphi_{1},\varphi_{2})&\;=\;1-2\varphi_{1}-2\varphi_{2}+\varphi_{1}^{2}-2\varphi_{1}\varphi_{2}+\varphi_{2}^{2}\;=\;\prod_{\epsilon_{i}=\pm1}(1+\epsilon_{1}\sqrt{\varphi_{1}}+\epsilon_{2}\sqrt{\varphi_{2}}),
\end{aligned}\end{equation}
which vanishes precisely when $\widetilde{M}$ degenerates into a single point. The fundamental period of $\widetilde{M}$ is
\begin{equation}\begin{aligned}
\varpi_{0}(\varphi_{1},\varphi_{2})&=\sum_{m_{i}\geq0}\left(\frac{(m_{1}+m_{2})!}{m_{1}!\;\;m_{2}!}\right)^{2}\varphi_{1}^{m_{1}}\varphi_{2}^{m_{2}}.
\end{aligned}\end{equation}
The infinite series is provided by the methods of \cite{Hosono:1994ax}. One can recognise the following two closed forms, using an Appell series $F_{4}$ and integrals of products of Bessel functions\footnote{Also known as a Bessel moment.} (for $\text{Re}[\sqrt{\varphi_{1}}]+\sqrt{\varphi_{2}}]<1$, so that the integral converges). 
\begin{equation}
\varpi_{0}(\varphi_{1},\varphi_{2})=F_{4}(1,1;1,1;\varphi_{1},\varphi_{2})=\int_{0}^{\infty}u\,K_{0}(u)I_{0}(u\sqrt{\varphi_{1}})I_{0}(u\sqrt{\varphi_{2}})\;\text{d}u.
\end{equation}

The mirror $M$ of $X$ is then obtained as the $\mathbb{Z}_{2}$ quotient of $\widetilde{M}$ by the action $U\leftrightarrow V$, after setting $\varphi_{1}=\varphi_{2}=\varphi$. This action is fixed-point free provided $\Delta(\varphi,\varphi)=1-4\varphi$ does not vanish. The fundamental period of $M$ is then
\begin{equation}
\varpi_{0}(\varphi)=\varpi_{0}(\varphi,\varphi)=\frac{1}{\sqrt{1-4\varphi}},
\end{equation}
in agreement with the GLSM result \eqref{eq:GLSMperiod}. Note that the pole of $\varpi_{0}$ at $\varphi=1/4$, where $\Delta=0$, was explained in the GLSM analysis of \sref{sect:CoulombBranches} as the Coulomb branch locus. For threefolds, such Coulomb branch loci are mirrors to conifold points (or their quotient generalisation, hyperconifold points \cite{Davies:2013pna}), where three-cycles on the mirror manifold collapse to nodal singularities. In superstring compactifications, branes wrapping these cycles give rise to massless hypermultiplets. The relation between massless hypermultiplets in string compactifications and Coulomb branches in the GLSM was identified in \cite[\S 6.1]{Hori:2011pd} . For our present example, the Coulomb branch singularity has the mirror interpretation as the locus in $\varphi$-space where the quotient group is not freely acting. From \eqref{eq:MirrorSolution}, one sees that for $\varphi_{1}=\varphi_{2}=\varphi=1/4$ the cover $\widetilde{M}$ is not two points, but instead is a single point $(U,V)=(\frac{1}{2},\frac{1}{2})$ that is mapped to itself by the $\mathbb{Z}_{2}$ quotient action $U\leftrightarrow V$. In the GLSM analysis, the Coulomb branch in $\sigma_{i}$-space for $\text{e}^{-t}=\frac{1}{4}$ is similarly fixed by the $\mathbb{Z}_{2}$ Weyl group action.

\section{Where to go from here?}
\label{sec-beyond}
The model we have discussed here, and non-regular GLSMs in general, leave many open questions and possibilities for further research. 
\begin{itemize}
\item {\bf Witten index.} One expects that phases of GLSMs have matching Witten indices.  At $\zeta\gg0$, the vacuum of the GLSM is a single point, while at $\zeta\ll0$ we find two points from the Higgs branch and a further point from the Coulomb branch. For non-Calabi-Yau GLSMs, where there can be additional massive Coulomb vacua in addition to Higgs vacua, the Witten index coming from the Higgs branch has to be corrected by adding the number of Coulomb vacua. If we did this for our model we would find a mismatch of $1$ vs.~$3$. We could achieve a matching if the extra Coulomb branch contributes with a negative sign, but we have no explanation why this should be the case. Computing the GLSM elliptic genus \cite{Benini:2013nda,Benini:2013xpa}  or finding a way to compute the Witten index directly in the $\zeta\ll0$ phase could help clarify this issue.
\item {\bf D-brane transport and categorical equivalences.} The B-brane categories of phases of GLSMs are expected to be equivalent. This is understood for abelian GLSMs \cite{Herbst:2008jq} and for some regular non-abelian GLSMs, including those for the R{\o}dland and Hosono-Takagi models \cite{MR3223878,MR3673174,grr}. It would be interesting to generalise these results to non-regular GLSMs. A recent result \cite{Guo:2025yed} makes a proposal for a prescription for D-brane transport in a non-regular model which provides valuable guidance.

Statements on categorical equivalences on our class of models, including non-regular ones, can be found in the mathematics literature \cite{MR4065183,HT2015-2,MR405345920200101}. For non-regular models, the statements of categorical equivalences include adding further exceptional collections, similar to what happens in the non-Calabi-Yau case where the exceptional collections can be related to the massive Coulomb vacua. However, in the case of non-regular models like ours, the mathematics result imply that the category associated to the $\zeta\gg0$ phase would have to be modified and not the category associated to the $\zeta\ll0$ phase, where our Coulomb branch appears. While we do not know how to understand this in the context of GLSMs, this may also give hints on what the accurate way to compute Witten indices for non-regular models is. Similar statements have also been made in the mathematics literature for non-regular models associated to Grassmannians \cite{kuznetsov06}. 

  \item {\bf Non-abelian duality} A duality for non-abelian GLSMs has been found in \cite{Hori:2011pd}. While non-regular models have been excluded from the discussion, our model is covered by the construction of the dual theories. In particular, the statement is that the non-abelian dual of a Hosono-Takagi-type GLSM with gauge group\footnote{We refer to the original paper for the difference between $O(k)_+$ and $O(k)_-$.} $O(k)_+$ with $N$ fundamentals is $SO(N-k+1)$. For our case of $N=k=2$, the dual non-abelian factor would be trivial and we would expect an abelian dual. It would be interesting to analyse this further and to find out if the duality requires modifications for non-regular GLSMs. There are also other interesting abelian models describing a single point which do not necessarily come from a non-abelian duality. See below for some examples. 

\item {\bf Non-abelian mirrors} Mirrors for non-abelian GLSMs have been proposed in \cite{Hori:2000kt, Gu:2018fpm}. It would be interesting to see if these constructions also work for non-regular GLSMs. 

\item{\bf Divergences} We have not provided a physical interpretation of \eqref{eq:ZetaLog}. In this work we have relied on analytic properties of ${}_{2}F_{1}$ that were not available to us in \cite{Knapp:2025hnf}. It remains to better understand the divergences of the series provided by GLSM techniques in non-regular models.

\item {\bf Other realisations of two points}
We have already mentioned the $U(1)$ model describing $\mathbb{P}^{1}[2]$, studied in \cite{Hori:2011pd}. This has three chiral superfields $(p,x_{0},x_{1})$ with $U(1)$ charges $(-2,1,1)$. At negative FI-parameter the $U(1)$ model flows to a Landau-Ginzburg orbifold.

Another way to realise two points is through a $U(2)$ GLSM with 4 chiral superfields $x_{i}$ that transform in the fundamental representation and four chiral superfields $p_{i}$ transforming in the $\text{det}^{-1}$ representation. Phase analysis at $\zeta>0$ reveals an intersection of four hyperplanes in $\text{Gr}(2,4)$. This model is non-regular. The fundamental period of the intersection $\text{Gr}(2,4)[1,1,1,1]$ can be obtained by using the ``trick with the factorials" from \cite{Batyrev:1998kx}, recovering $\varpi_{0}(\varphi)=\frac{1}{\sqrt{1-\varphi}}$.
  
  \item {\bf Other realisations of one point} 
  Motivated by the application in \cite{Hori:2006dk} of a simpler dual abelian description to study an non-regular (threefold) model, we describe some other ways that a GLSM can realise a single point, which may help study our main model.
  
 The $U(1)$ model describing $\mathbb{P}^{1}[2]$ can be supplemented by an additional gauging, for suitable superpotential. We can replace the gauge group by $U(1)\times\mathbb{Z}_{2}$, where the $\mathbb{Z}_{2}$ acts trivially on the single $p$-field and exchanges $x_{1}\leftrightarrow x_{2}$, whereupon the $\zeta\gg0$ phase becomes a single point. We expect in this $U(1)\times\mathbb{Z}_{2}$ model to find a further Landau-Ginzburg orbifold in the $\zeta<0$ phase.

Furthermore, we have discussed in the main body of our article the $(U(1)\times U(1))\rtimes\mathbb{Z}_{2}$ model with $U(1)\times U(1)$ charge matrix given in \eqref{eq:ChargeMatrix}. One could consider the same charge matrix, with two unequal FI parameters $\zeta_{1},\zeta_{2}$ for an abelian model with gauge group $U(1)^{2}\times\mathbb{Z}_{2}$. Here the $\mathbb{Z}_{2}$ action has generator $p_{1}\leftrightarrow p_{2}$, $x_{1}\leftrightarrow x_{2}$, $y_{1}\leftrightarrow y_{2}$. For suitable superpotential the vacuum manifold at $\zeta_{1}>0,\zeta_{2}>0$ is a freely acting quotient of two points, again giving a point. 
  \end{itemize}

\begin{acknowledgement}
We thank the participants of the MATRIX program ``The geometry of moduli spaces in string theory'' for creating a wonderful atmosphere that sparked many interesting discussions, including one on Calabi-Yau zerofolds. We thank Kentaro Hori, Shinobu Hosono, and Hiromichi Takagi for patiently explaining their work to us. We also would like to congratulate MATRIX Institute on the occasion of its tenth anniversary and wish many happy returns. Finally, thanks to the editors of the proceedings volume, Jock McOrist and S{\'e}bastien Picard, for accepting our contribution at the very last minute. JK is supported by the Australian Research Council Future Fellowship FT210100514. JM is supported by a University of Melbourne Establishment Grant.
\end{acknowledgement}
%


\providecommand{\href}[2]{#2}\begingroup\raggedright\endgroup

\end{document}